\documentclass[sigconf]{acmart}
\AtBeginDocument{%
  }

\setcopyright{acmlicensed}
\copyrightyear{2025}
\acmYear{2025}
\acmDOI{XXXXXXX.XXXXXXX}
\acmConference[CIKM '25]{Proceedings of the 35th ACM Conference on The Conference on Information and Knowledge Management}{November 10--14,
  2025}{Seoul, Korea}
\acmISBN{978-1-4503-XXXX-X/2018/06}

\usepackage{array}
\newcolumntype{P}[1]{>{\raggedright\arraybackslash}p{#1}}

\usepackage{multirow}
\usepackage{makecell} 
\usepackage{tikz}
\usepackage{pgfplots}
\usepackage{cleveref}
\pgfplotsset{compat=1.18}
\usepackage[compact]{titlesec}
\titlespacing{\section}{0pt}{*1}{*0.5}
\titlespacing{\subsection}{0pt}{*1}{*0.5}
\setcopyright{none}              
\acmConference{}{}{}             
\acmYear{}                       
\acmISBN{}                       
\acmDOI{}                        
\acmPrice{}                      

\makeatletter
\renewcommand\@formatdoi[1]{}
\renewcommand\@copyrightpermission{}
\renewcommand\footnotetextcopyrightpermission[1]{}
\makeatother

\begin{document}

\title{LLM4ES: Learning User Embeddings from Event Sequences via Large Language Models}

\author{Aleksei Shestov}
\affiliation{
	\institution{Sber AI Lab}
	\city{Moscow}
	\country{Russian Federation}
}
\orcid{0009-0005-5207-7610}
\email{shestovmsu@gmail.com}

\author{Omar Zoloev}
\affiliation{
	\institution{Sber AI Lab}
	\city{Moscow}
	\country{Russian Federation}
}
\affiliation{
	\institution{NUST MISIS}
	\city{Moscow}
	\country{Russia}
}
\orcid{0009-0004-9447-5599}
\email{ozoloevwork@gmail.com}

\author{Maksim Makarenko}
\orcid{0000-0002-6761-8695}
\email{molegmakarenko@sberbank.ru}
\affiliation{
	\institution{Sber AI Lab}
	\city{Moscow}
	\country{Russian Federation}
}

\author{Mikhail Orlov}
\orcid{0009-0007-1978-9995}
\email{ormian@mail.ru}

\author{Egor Fadeev}
\orcid{0009-0005-2111-6760}
\email{edfadeev@sberbank.ru}

\author{Ivan Kireev}
\orcid{0009-0004-1618-8981}
\email{ivkireev@yandex.ru}
\affiliation{
	\institution{Sber AI Lab}
	\city{Moscow}
	\country{Russian Federation}
}
\author{Andrey Savchenko}
\orcid{0000-0001-6196-0564}
\email{avsavchenko@hse.ru}
\affiliation{
	\institution{Sber AI Lab}
	\city{Moscow}
	\country{Russian Federation}
}
\affiliation{
	\institution{HSE University, Laboratory of Algorithms and Technologies for Networks Analysis}
	\city{Nizhny Novgorod}
	\country{Russian Federation}
}


\renewcommand{\shortauthors}{Shestov et al.}

\begin{abstract}
This paper presents LLM4ES, a novel framework that exploits large pre-trained language models (LLMs) to derive user embeddings from event sequences. Event sequences are transformed into a textual representation, which is subsequently used to fine-tune an LLM through next-token prediction to generate high-quality embeddings. We introduce a text enrichment technique that enhances LLM adaptation to event sequence data, improving representation quality for low-variability domains. Experimental results demonstrate that LLM4ES achieves state-of-the-art performance in user classification tasks in financial and other domains, outperforming existing embedding methods.
The resulting user embeddings can be incorporated into 
a wide range of applications, from user segmentation in finance to patient outcome prediction in healthcare.
\end{abstract}



\maketitle
\vspace{-5pt}
\section{Introduction}
Event-based sequential (ES) data prevails across many domains, including e-commerce \cite{Dai2023}, healthcare \cite{Das2023}, education \cite{Liu2023XES3G5M}, and finance \cite{babaev2022coles}. This data type captures interactions and actions unfolding over time - such as transactions, medical procedures, user clicks, or learning activities — and is foundational for a wide range of applications, including user modeling and recommender systems \cite{Petrov2023}, anomaly detection \cite{Elmougy2023}, and personalized prediction tasks such as knowledge tracing in education \cite{Liu2023SimpleKT}. 

\begin{figure}[t]
    \centering
    \includegraphics[width=\linewidth]{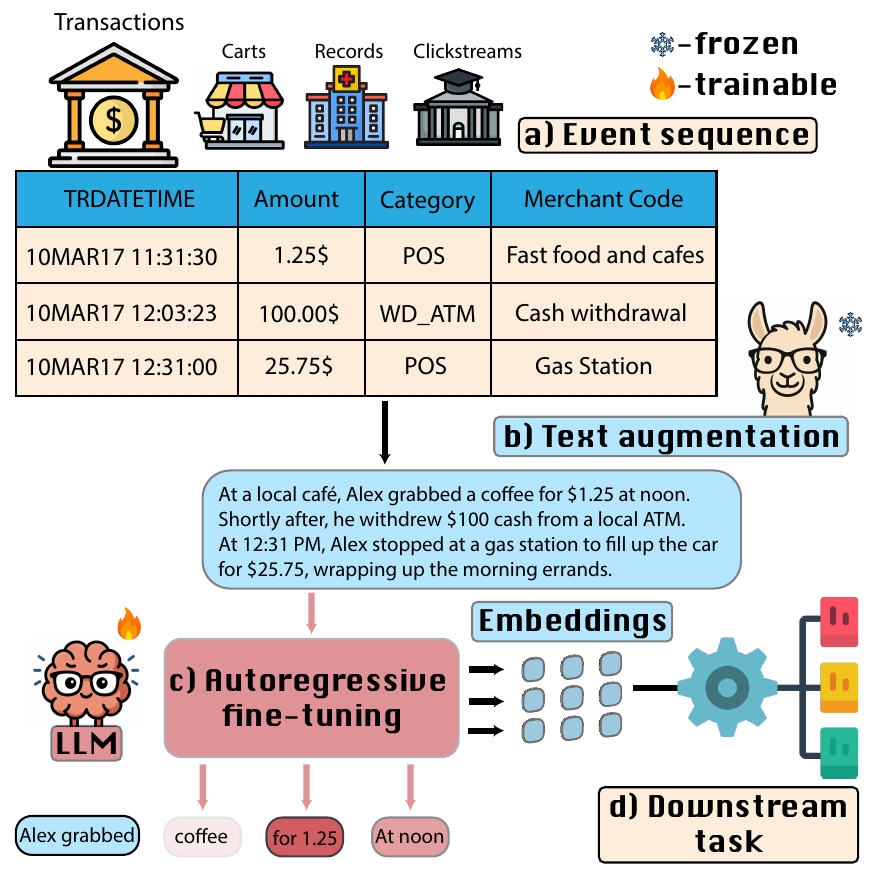}
    \caption{The proposed LLM4ES framework. (a) The event stream is converted into text. (b) A pre-trained LLaMA model generates enriched event descriptions. (c) LLM4ES is fine-tuned on the enriched data using an autoregressive objective. (d) The resulting embeddings are used for downstream tasks.}
    \label{fig:embedding_pipeline}
\end{figure}

A core challenge in leveraging ES data is the effective construction of user representations. Embedding-based approaches offer a powerful solution by mapping sequences into fixed-length vector representations that can be used across diverse tasks~\cite{gui2024survey}. Typical methods for constructing user embeddings from sequential event data often rely on self-supervised learning techniques such as contrastive learning \cite{babaev2019rnn}, next event prediction \cite{skalski2023towards}, and Contrastive Predictive Coding (CPC) \cite{oord2018representation}. 

With recent advancements in large language models (LLMs), there is an opportunity to improve upon these methods. LLMs, trained on large amounts of data, encode rich knowledge about event semantics and behavior patterns that cannot be directly derived from event sequence datasets alone. However, key challenge in applying LLMs to structured event sequence data lies in the distributional shift between their original textual training data and domain-specific formats~\cite{suntest, abdullaeva2024esqa}. Existing adaptation approaches across domains, including time series analysis~\cite{suntest,jintime}, tabular data processing~\cite{hegselmann2023tabllm}, recommender systems~\cite{bao2023tallrec,zheng2024harnessing,yin2023heterogeneous}, and financial transaction analysis~\cite{abdullaeva2024esqa}, typically focus on supervised fine-tuning or instruction tuning for specific downstream tasks. 

In contrast, this work adapts LLMs to extract general-purpose user embeddings from event sequences broadly applicable beyond predefined tasks. In particular, 
we propose \texttt{LLM4ES}\footnote{The source code to reproduce experiments is available at \url{https://anonymous.4open.science/r/LLM4ES-0385/}.} (Figure~\ref{fig:embedding_pipeline}), a framework that fine-tunes LLMs on enriched textual representations of user event sequences. This approach enables the extraction of semantically rich user embeddings suitable for various downstream applications.
We summarize our contributions as follows: 
\begin{enumerate} 
\item We introduce LLM4ES, a novel framework for extracting user embeddings by fine-tuning LLMs on domain-specific event sequence data converted into an enriched text-based format. LLM4ES achieves up to 7\% relative improvement in ROC-AUC both on financial and non-financial datasets over SOTA embedding methods, both individually and when used in an ensemble with other methods such as CoLES\cite{babaev2022coles}. 
\item We propose a new general technique for adapting LLMs to event sequence data by text enrichment, enabling effective training on the low-variability domain. This method yields up to a 1.8\% relative improvement in ROC-AUC on financial and non-financial datasets compared to training on raw serialized data.
\end{enumerate}

\section{Proposed approach}
Figure~\ref{fig:embedding_pipeline} presents an overview of the proposed approach. First, event sequence data are converted to a text-based format with human-readable field descriptions (Fig.~\ref{fig:embedding_pipeline}a). To enhance knowledge extraction, we apply a pre-trained LLM to transform event sequences (Fig.~\ref{fig:embedding_pipeline}b) into a richer textual format. The language model is fine-tuned on enriched data using a next-token prediction objective, and embeddings are obtained by averaging the representations from the final hidden layers (Fig.~\ref{fig:embedding_pipeline}c). We evaluate LLM4ES embeddings on open-source transactional datasets, where it consistently outperforms state-of-the-art (SOTA) methods~\cite{skalski2023towards, babaev2022coles, karpukhin2024DeTPP} and strong baselines~\cite{clark2020electra, oord2018representation} (Fig.~\ref{fig:embedding_pipeline}d). 
We describe each step in more detail in the following subsections.

\textbf{Event sequence serialization.} To process structured event sequence data with LLMs, we convert the event history of each user into a text-based format, in which the first line serves as a header with feature names, separated by the ``|'' delimiter. Subsequent lines represent individual events, with feature values separated by commas. Ordinal features, such as merchant category code (MCC), are replaced with their corresponding textual descriptions (e.g., MCC=5411 becomes ``Grocery Stores, Supermarkets''). 



\begin{table}[h]
    \caption{Sample prompt for ES data transformation. }
    \label{tab:prompt_example}
\vspace{-10pt}
\resizebox{\columnwidth}{!}{
    \begin{tabular}{|l|}
        \hline
        \textbf{System:} You are a specialized AI assistant designed for text data transformation. \\
        Your task is to take a structured list of financial transactions and convert them \\ into different text-based formats based on the user's request. \\
        Ensure clarity, accuracy, and proper formatting while preserving all original data fields. \\
        You are capable of transforming data into formats including: \\ 
        - JSON, Markdown, HTML, Plain text, tables.
        \\
        Always ensure the output is well-structured and visually clear. \\
        If the user specifies a custom format, follow the instructions closely. \\ Ask for clarifications when the format is ambiguous. \\
        \hline
        \textbf{User:} I have a list of financial transactions in a structured tabular format. \\
        Please creatively transform this data into a different textual representation. \\
        You can choose the format freely (JSON, HTML, Markdown, plain text, XML, YAML, \\ or anything else). \\
        \textbf{Requirements:} \\
        - Randomize the text format and structure creatively. \\
        - Paraphrase category names and transaction types while keeping the meaning clear. \\
        - Ensure the data remains understandable but presented differently. \\
        Generate the transformed output. \\
        \hline
    \end{tabular}
    }
\end{table}

\begin{table}
    \caption{Example of ES data transformed using the proposed text enrichment method}
    \vspace{-10pt}
    \label{tab:text_aug_example}
\small
    \begin{tabular}{|p{8cm}|}
        \hline
\textbf{Transaction Breakdown}\\
\textit{Categories and Subcategories}\\
\textit{Grocery Stores}
\begin{itemize}
    \item Supermarkets: 24 transactions, totaling 166 rubles
    \item Various Grocery Stores: 15 transactions, totaling 73 rubles
    \item Fast Food: 7 transactions, totaling 36 rubles
\end{itemize}
\textit{Service Stations}
\begin{itemize}
    \item With or without additional services: 7 transactions, totaling 33 rubles
\end{itemize}

{[{\Large$\cdot\cdot\cdot$}]\par}

\textbf{Transaction Details}

\small
\begin{tabular}{|P{0.8cm}|P{1.cm}|P{2.4cm}|P{2.2cm}|}
\hline
Date (UNIX) & Amount (rubles) & Merchant Category & Transaction Type \\
\hline
17298 & 4 & Grocery Stores & Point of Sale \\
17299 & 8 & Financial Institutions & Deposit \\
\hline
\multicolumn{4}{|c|}{\Huge$\cdot\cdot\cdot$} 
\\  
\hline
\end{tabular}

\vspace{0.3cm}  
\\
\hline
\end{tabular}  

\end{table}

\textbf{Text Enrichment.} To overcome the limited knowledge extraction caused by training on low-variability text, we propose a text enrichment technique. Specifically, we employ the \textbf{Llama-3.1-8B-instruct} model~\cite{dubey2024llama} to rewrite event sequence texts into diverse formats and styles. The prompt we utilize for this process is shown in Table~\ref{tab:prompt_example}, and Table~\ref{tab:text_aug_example} provides an example of a transformed user history format.

\textbf{LLM Fine-tuning.} We perform LLM fine-tuning on event sequences $\{\mathbf{x}_t\}$ that have been converted into text-enriched format (Table~\ref{tab:text_aug_example}). During fine-tuning, all weights of a pre-trained LLM are updated to minimize the conventional next-token prediction loss:
\begin{equation}
\mathcal{L}_{\text{NTP}} = - \sum_{t} \log P(x_t \mid \mathbf{x}_{<t}; \theta)
\end{equation}
Here, \( x_t \) denotes the token at position \( t \), \( \mathbf{x}_{<t} \) represents the sequence of preceding tokens, and \( \theta \) refers to the model parameters. 

\textbf{Embeddings Extraction.} We follow recent works advocating mean pooling over EOS pooling for LLM embeddings~\cite{behnamghader2024llmvec,muennighoff2024generative} and averaging of $k > 1$  last layers l~\cite{tang2024pooling} . In particular, we average hidden states $\mathbf{H}_l$ of all sequence tokens over the $k=8$ last hidden layers of the fine-tuned LLM to obtain a single fixed-size user embedding:

\begin{equation}
\mathbf{User}_{emb} = \text{MeanPooling} \left( \frac{1}{k} \sum_{l=L-k}^{L - 1} \mathbf{H}_l \right),
\end{equation}
where \( L \) is the total number of layers in the LLM. 


\textit{Datasets.}
We evaluate LLM4ES on diverse datasets spanning financial and non-financial domains, including proprietary and public sources. The primary set of experiments focuses on financial transaction data, while additional benchmarks assess cross-domain generalization and real industrial applicability.\\
\textbf{Public financial datasets (main experiments)}: The core evaluation is based on three publicly available banking transaction datasets from ML competitions, each associated with a distinct user classification task (Table~\ref{tab:dataset_summary}). These datasets contain sequences of debit card transactions with both labeled and unlabeled samples and serve as the main testbed for developing and analyzing LLM4ES.\\
\textbf{Internal corporate dataset}: To assess real-world deployment scenarios, we include a proprietary financial dataset focused on customer lifetime value (CLTV) prediction. This dataset features long-term user transaction histories with continuous targets.\\
\textbf{Non-financial benchmark}: To evaluate the model’s generalization across domains, we include a publicly available dataset from a non-financial context: MovieLens 100k (user preferences)\cite{harper2015movielens}. This dataset differs significantly in structure and semantics from financial data, allowing us to test the transferability of LLM4ES to other types of event sequences. It contains preference sequences for 943 users on 1682 movies, where each event includes a movie title, year of production, and the user-assigned rating score. We evaluate pre-trained user embeddings on downstream gender and age classification tasks.
\vspace{-5pt}
\begin{table}[h!]
\centering
\caption{Financial datasets used in the experimental study}
\vspace{-10pt}
\label{tab:dataset_summary}
\resizebox{\columnwidth}{!}{ 
\begin{tabular}{|c|c|c|c|c|c|}
\hline
Dataset & Labeled Samples & Unlabeled Samples & Num. Features & Target \\ 
\hline
Rosbank~\cite{rosbank} & 5,217 & 5,000 & 5 & churn \\
Gender~\cite{gender} & 5,000 & 10,000 & 5 & gender \\
Age~\cite{agepred} & 30,000 & 20,000 & 4 & age group \\
\hline
\end{tabular}
}
\end{table}

\textit{Validation Strategy.}
For each dataset, we reserved 10\% of the labeled users as a test set for evaluation. The remaining 90\% of labeled data, combined with all available unlabeled data, formed the training set. To evaluate the quality of representations, we implemented the following 5-fold cross-validation procedure. The training set was partitioned into five approximately equal-sized disjoint subsets (folds). 
For each fold $v$, 1) a LightGBM\cite{guolin2017lightgbm} classifier was trained on the embeddings from the remaining four folds, and 2) the trained LightGBM classifier was evaluated on the test set embeddings, yielding a performance metric $M_v$. We compute the mean ($\mu_M$) and standard deviation ($\sigma_M$) of ROC-AUC across all folds. The final reported performance is $\mu_M \pm \sigma_M$, providing an estimate of both the expected ROC-AUC and its stability across different data partitions.

\textit{Baselines.}
We compared LLM4ES against SOTA methods and strong baselines from various domains:\\
\textbf{Event sequences:} CoLES~\cite{babaev2022coles} – Pairwise contrastive learning for event sequences; 
NPPR~\cite{skalski2023towards} – Embedding training via bidirectional prediction of past and future events.\\
\textbf{Temporal point processes (TPP):} DeTPP~\cite{karpukhin2024DeTPP} – Captures event timing and type distributions; 
IFTPP~\cite{shchurintensity} – MAE for time prediction and cross-entropy for event classification, with GRU/Transformer backbones.\\
\textbf{Natural language processing:} NSP~\cite{devlin2019bert} – BERT’s next sentence prediction; 
RTD~\cite{clark2020electra} – Replaced token detection with 15\% event substitutions.\\
\textbf{Tabular data:} Aggregated features – mean, std, min, max for numeric fields (e.g., amount), and grouped statistics for categorical fields (e.g., MCC).

For the TPP baselines, their training procedure was employed as a pretraining objective. After pretraining, the sequence embedding was obtained by either using the final hidden state (GRU) or the embedding of the last token (Transformers). All baseline methods were implemented with their recommended optimal settings as described in their respective papers.

\textit{LLM4ES Setup Details.}
In all experiments, we use the LLama-3.2-3B \cite{meta2024llama} as a base model for LLM4ES. 
The base model was fine-tuned for 30 epochs with a learning rate of $1 \times 10^{-5}$, using a cosine learning rate scheduler and a weight decay of $1 \times 10^{-6}$. We used a batch size of 1, mixed precision training (bf16), and an Adam optimizer with $\beta_2 = 0.95$, $\epsilon = 1 \times 10^{-5}$. For each method, optimal epoch selection uses the mean ROC-AUC 
computed across training folds to identify the best checkpoint.
\vspace{-2pt}
\section{Experimental results}
\subsection{Comparison with other methods}
\subsubsection{Benchmarking on financial transaction data.}
Table~\ref{tab:dataset_metrics} reports ROC-AUC scores across three datasets. LLM4ES achieves the highest performance on both the Rosbank and Age datasets, with relative improvements of +0.7\% over NPPR on Rosbank and +2.3\% over CoLES on Age. On the Gender dataset, it remains highly competitive, scoring 87.5\% compared to 88.2\% for CoLES. Notably, LLM4ES delivers a significant relative gain of up to 10\% on the Age dataset (+9.6\% over CPC, +9.6\% over Agg, and +6.1\% over NSP), specifically +6.3\% over the transformer-based IFTPP-T. Thus, LLM4ES consistently outperforms specialized architectures designed for domain-specific event sequences modeling. 
\vspace{-5pt}
\begin{table}[H]
\caption{Main results for financial datasets, ROC-AUC.}
\vspace{-10pt}
\label{tab:dataset_metrics}
\small
\begin{tabular}{|c|c|c|c|}
\hline
Method & Rosbank   & Gender  & Age \\ 
\hline
agg & 0.827 $\pm$ 0.010 & \underline{0.877 $\pm$ 0.004} & 0.629 $\pm$ 0.002 \\
\hline
CPC & 0.792 $\pm$ 0.015 & 0.851 $\pm$ 0.006 & 0.602 $\pm$ 0.004 \\
\hline
RTD & 0.771 $\pm$ 0.016 & 0.855  $\pm$  0.008 & 0.631  $\pm$ 0.006 \\
\hline
CoLES & 0.841 $\pm$ 0.005 & \textbf{0.882 $\pm$ 0.004} & \underline{0.644 $\pm$ 0.005} \\ 
\hline
NSP & 0.828 $\pm$ 0.012& 0.852 $\pm$ 0.011 & 0.621 $\pm$ 0.005 \\
\hline
NPPR & \underline{\(0.845 \pm 0.003\)} & -  & \(0.642 \pm 0.001\) \\ \hline
DeTPP & 0.823 & - & 0.632 \\
\hline
IFTPP & 0.828 $\pm$ 0.004 & 0.863 $\pm$ 0.003 & 0.632 $\pm$ 0.003 \\
\hline
IFTPP-T & 0.814 $\pm$ 0.004 & 0.852 $\pm$ 0.005 & 0.620 $\pm$ 0.002 \\
\hline
\textbf{LLM4ES} & \textbf{0.851 $\pm$ 0.004} & 0.875 $\pm$ 0.004 &  \textbf{0.659 $\pm$ 0.004} \\
\hline
\end{tabular}

\end{table}

\subsubsection{Ensemble Performance.}

We evaluated the performance of LLM4ES as part of an ensemble formed by concatenating embeddings. As shown in Table~\ref{tab:ensemble_metrics}, the combination of LLM4ES with CoLES consistently outperforms standalone methods—resulting in +1.1\% relative improvement on gender prediction and +3.3\% gain on age prediction compared to standalone CoLES. Notably, the CoLES + Agg ensemble improves gender prediction by +0.9\% over standalone CoLES but reduces performance on the age prediction task by -2.3\%. In addition, the CoLES+LLM4ES ensemble achieves up to +4.2\% relative improvement over CoLES+IFTPP and +4.7\% over CoLES+IFTPP-T on the age dataset. These results indicate that LLM4ES captures complementary aspects of user behavior that are not fully exploited by existing methods.

\begin{table}[H]
\caption{ROC-AUC of ensembles (concatenated embeddings)
}
\vspace{-10pt}
\label{tab:ensemble_metrics}
\small
\begin{tabular}{|c|c|c|c|}
\hline
Method & Gender & Age \\ 
\hline
CoLES + agg  & 0.890$\pm$ 0.004 & 0.629 $\pm$ 0.002 \\
\hline
CoLES + IFTPP & 0.884 $\pm$ 0.002 & 0.638 $\pm$ 0.005 \\
\hline
CoLES + IFTPP-T & 0.880 $\pm$ 0.004 & 0.635 $\pm$ 0.004 \\
\hline
CoLES + LLM4ES & \textbf{0.892 $\pm$ 0.004} & \textbf{0.665 $\pm$ 0.004} \\ 
\hline
\end{tabular}
\end{table}

\subsubsection{Broader domains.}

In this section, we evaluate the generalization ability of LLM4ES on a broader set of data sources beyond the financial domain. \Cref{tab:domain_ablation} summarizes the obtained results. On MovieLens, the model improves gender classification ROC-AUC by 9.5\% (from 0.683 to 0.748) and reduces age prediction MAE by 15.6\% (from 8.70 to 7.34), clearly outperforming the best-performing in other settings CoLES baseline. 

\begin{table}[h]
\centering
\small
\caption{Cross-domain evaluation on non-financial (MovieLens) and financial (CLTV) tasks}
\vspace{-10pt}
\label{tab:domain_ablation}
\begin{tabular}{|l|c|c|c|}
\hline
Model & \multicolumn{2}{c|}{MovieLens 100k} & CLTV Prediction \\
\cline{2-3}
      & Gender (Accuracy) & Age (MAE) & MAE \\
\hline
CPC       & 0.567 & 9.97 & - \\
\hline
CoLES     & 0.683 & 8.70 & $1.6 \times 10^4$ \\
\hline
LLM4ES    & \textbf{0.748} & \textbf{7.34} & \textbf{1.56} $\times 10^4$ \\
\hline
\end{tabular}
\end{table}

We also evaluate LLM4ES on a proprietary internal financial dataset for customer lifetime value (CLTV) prediction. The model achieves a lower prediction error compared to the CoLES baseline, improving performance from $1.6 \times 10^4$ to $1.56 \times 10^4$ MAE.

\subsection{Ablation studies and component analysis}
\subsubsection{Model component analysis.}
To evaluate the impact of each component in LLM4ES, we conducted ablation studies summarized in Table~\ref{tab:text_augment}. Fine-tuning a pre-trained LLM on serialized transactions improves ROC-AUC from 82.3\% to 84.0\% on Rosbank and from 71.1\%  to 85.0\% on Gender. Replacement of raw sequences with text-enriched inputs further boosts performance to 84.9\% and 86.7\%, respectively. The best results—85.1\% on Rosbank and 87.5\% on Gender—are achieved when training on enriched data from both datasets. These results confirm the effectiveness of both fine-tuning and text enrichment in the proposed framework.

\vspace{-3pt}
\begin{table}[H]
\caption{Ablation study of the LLM4ES components}
\vspace{-10pt}
\label{tab:text_augment}
\resizebox{\columnwidth}{!}{
\begin{tabular}{ |c|c|c|c|}
\hline
LLM & Fine-tuning Data & Rosbank & Gender \\
\hline
Pre-trained & - & 0.823 $\pm$ 0.003 & 0.711 $\pm$ 0.007 \\
\hline
& Original serialized & 0.839 $\pm$ 0.003 & 0.850 $\pm$ 0.003 \\
Fine-tuned & Text enriched & \underline{0.849 $\pm$ 0.001} & \underline{0.867 $\pm$ 0.004} \\
& \makecell[l]{Text enriched data\\ from both datasets} & \textbf{0.851 $\pm$ 0.004} & \textbf{0.875 $\pm$ 0.004} \\
\hline
\end{tabular}
}
\end{table}

\subsubsection{LLM architecture ablation.}
\label{sec:llm-ablation}

We evaluate the text enrichment method on Qwen-3-1.7B~\cite{qwen3technicalreport} to assess its generalization across architectures. As shown in \Cref{tab:llm_ablation}, it improves ROC-AUC by approximately 1\% for both Qwen and Llama-3.2-3B, demonstrating effectiveness beyond the primary model.

\begin{table}[h]
\caption{LLM type ablation, Rosbank dataset}
\vspace{-10pt}
\label{tab:llm_ablation}
\centering
\small
\begin{tabular}{|l|c|c|}
\hline
Base LLM & Data representation & ROC-AUC \\
\hline
Qwen 3 1.7b & Original Serialized & 0.834 \\
\hline
Qwen 3 1.7b & Text-enriched & \underline{0.844} \\
\hline
LLama 3.2 3b & Original serialized & 0.839 $\pm$ 0.003 \\
\hline
LLama 3.2 3b & Text-enriched & \textbf{0.849 $\pm$ 0.001} \\
\hline
\end{tabular}
\end{table}

\subsubsection{Text enrichment analysis.}

Table~\ref{tab:augmentation_formats} shows that mixed-format text enrichment (0.849) surpasses the best single-format (0.845) at equal data size, confirming benefits of format diversity. Increasing enriched data volume to 7x further improves performance (0.854), demonstrating its role as effective data augmentation. Adding raw serialized data degrades results (0.843), likely because enrichment provides clearer transaction summaries aiding semantic understanding. Compared to HKFR  \cite{yin2023heterogeneous} and TALLRec \cite{bao2023tallrec}, HKFR text representation yield lower scores (0.823 train-only, 0.756 train+test), while TallRec’s raw serialization (0.839) performs better but remains inferior to LLM4ES. These results confirm the superior effectiveness of the proposed text enrichment approach.

\begin{table}[h]
\caption{Text Enrichment Ablation, Rosbank Dataset}
\vspace{-10pt}
\label{tab:augmentation_formats}
\centering
\small
\begin{tabular}{|l|c|c|}
\hline
Method & Data Amount & ROC-AUC \\
\hline
Raw serialization, TallRec~\cite{bao2023tallrec} & 1x & 0.839 $\pm$ 0.003 \\
\hline
Best single format (HTML) & 1x & 0.845 $\pm$ 0.002 \\
\hline
Worst single format (JSON) & 1x & 0.839 $\pm$ 0.006 \\
\hline
Mixed formats & 1x & \underline{0.849 $\pm$ 0.001} \\
\hline
Mixed formats & 7x & \textbf{0.854 $\pm$ 0.004} \\
\hline
Mixed + raw & 8x (7x+1x) & 0.843 $\pm$ 0.003 \\
\hline
HKFR~\cite{yin2023heterogeneous}, only train & 1x & 0.823 $\pm$ 0.006 \\
\hline
HKFR~\cite{yin2023heterogeneous}, train \& test & 1x & 0.756 $\pm$ 0.008 \\
\hline
\end{tabular}
\end{table}

\vspace{-5pt}
\subsubsection{Training regime ablation.}

To assess the impact of different fine-tuning strategies, we compare full model fine-tuning (used in main approach) with parameter-efficient fine-tuning using LoRA~\cite{hulora} on the Rosbank dataset. While LoRA offers benefits in memory efficiency and training speed, it resulted in a lower ROC-AUC of 0.830±0.003. In contrast, full model fine-tuning achieved a superior ROC-AUC of 0.849±0.001, highlighting its advantage in terms of predictive performance on small-scale transactional data.

\subsubsection{Data size ablation.}
\begin{figure}[htbp] 
\centering
\small
\caption{Data size ablation, Rosbank dataset}
\label{pic:few_shot}
\begin{tikzpicture}
\begin{axis}[
    width=\columnwidth,   
    height=0.45\columnwidth, 
    xmode=log,
    xlabel={Sample size},
    ylabel={ROC-AUC},
    x label style={at={(axis description cs:0.5,-0.12)}, anchor=north}, 
    y label style={at={(axis description cs:-0.09,0.5)}, anchor=south},   
    xtick={1,10,100,500,1000,5000,10000},
    xticklabel style={rotate=45, anchor=east},
    grid=major,
    legend style={at={(0.95,0.05)}, anchor=south east},
    log ticks with fixed point,
    xmin=1, xmax=10000,
    ymin=0.7, ymax=0.9,
]

\addplot coordinates {
    (1,0.712) (10,0.733) (100,0.770) (500,0.824) (1000,0.828) (5000,0.830) (9717,0.841)
};
\addlegendentry{Coles}


\addplot coordinates {
    (1,0.820) (10,0.819) (100,0.828) (500,0.826) (1000,0.835) (5000,0.843) (9717,0.850)
};
\addlegendentry{LLM4ES}

\end{axis}
\end{tikzpicture}
\end{figure}
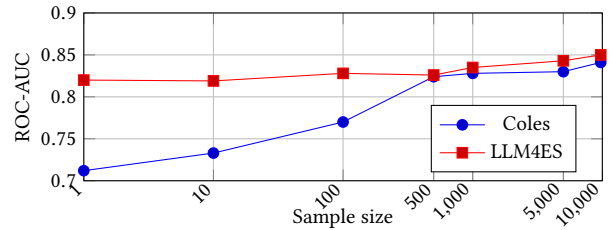

To evaluate robustness across varying data availability, we compare LLM4ES and CoLES over increasing sample sizes on the Rosbank dataset (Fig. \ref{pic:few_shot}). LLM4ES consistently outperforms CoLES, achieving ROC-AUC of 0.820 vs. 0.712 (+15.2\%) at 1 sample and 0.819 vs. 0.733 (+8.6\%) at 10 samples. Although the gap narrows with more data, it remains significant at full scale (0.850 vs. 0.841), confirming LLM4ES’s advantage across data regimes.

\vspace{-15pt}
\vspace{-20pt}
\section{Conclusion}

In this article, we presented LLM4ES, a novel approach for extracting user embeddings from event sequence data (Figure~\ref{fig:embedding_pipeline}). By converting event sequences into text and fine-tuning large language models (LLMs) on the resulting representations, we achieved significant improvements on downstream tasks such as user classification. The method reached state-of-the-art performance both as a standalone model (Table~\ref{tab:dataset_metrics}) and as part of an embedding ensemble (Table~\ref{tab:ensemble_metrics}). We also introduced a text enrichment technique that facilitates effective LLM training on low-variability sequences—a common challenge in real-world applications. A comprehensive ablation study highlights the contributions of text enrichment, full fine-tuning, and cross-domain training. These results show strong generalization to non-financial domains, with enriched text consistently improving model performance.




\section*{GenAI Usage Disclosure}
The core method development in this research does not involve LLMs as any important, original, or non-standard components. We used LLMs only to improve the quality of the main text.

\section*{Acknowledgments}
The work of A. Savchenko was prepared within the framework of the Basic Research Program at the National Research University Higher School of Economics (HSE).
\bibliographystyle{ACM-Reference-Format}
\bibliography{sample-base}

\end{document}